\documentclass[aps, prl, a4paper, twocolumn, 10pt, showpacs]{revtex4-1}
\usepackage{amsmath, amssymb, amsthm, bm, textcomp}
\usepackage[T1]{fontenc}
\usepackage[latin9]{inputenc}
\usepackage{graphicx,graphics}
\usepackage{geometry}
\usepackage{color}
\newcommand{\ket}[1]{\left|{#1}\right\rangle}
\newcommand{\bra}[1]{\left\langle{#1}\right|}

\newcommand{\be}{\begin{equation}}
\newcommand{\ee}{\end{equation}}
\newcommand{\eea}{\end{eqnarray}}
\newcommand{\bea}{\begin{eqnarray}}

\begin{document}

\title{Universal and optimal error thresholds for measurement-based entanglement purification}
%\title{Optimal universal error thresholds for measurement-based entanglement purification}

\author{M. Zwerger$^1$, H. J. Briegel$^{1,2}$ and W.\ D\"ur$^1$}

\affiliation{$^1$ Institut f\"ur Theoretische Physik, Universit\"at Innsbruck, Technikerstr. 25, A-6020 Innsbruck,  Austria.\\
  $^2$Institut f\"ur Quantenoptik und Quanteninformation der \"Osterreichischen Akademie der Wissenschaften, Innsbruck, Austria}
\date{\today}

\begin{abstract}
We investigate measurement-based entanglement purification protocols (EPP) in the presence of local noise and imperfections. We derive a universal, protocol-independent threshold for the required quality of the local resource states, where we show that local noise per particle of up to 24\% is tolerable. This corresponds to an increase of the noise threshold by almost an order of magnitude, based on the joint measurement-based implementation of sequential rounds of few-particle EPP. We generalize our results to multipartite EPP, where we encounter similarly high error thresholds.
\end{abstract}
%
%\pacs{03.67.Lx}
%%75.10.Hk    Classical spin models
%%75.10.Pq    Spin chain models
%%02.70.-c    Computational techniques
%%03.67.Lx    Quantum computation

\maketitle

% ---------------------------------------------------
% Introduction
% ---------------------------------------------------

%\section{Introduction}
\textit{Introduction.---}
Quantum entanglement shared between spatially separated parties is the key resource for many applications in quantum communication and computation. The generation of high-fidelity entangled states between parties connected by noisy quantum channels is hence a crucial step towards the implementation of quantum technologies. Possible applications range from quantum teleportation over quantum cryptography to entanglement-based implementation of quantum gates in distributed quantum computation. Quantum error correction \cite{Sho95,Ste96} as well as entanglement purification \cite{Be96,De96,Du07} offer a way to achieve this task, even in the presence of local noise and imperfections. Entanglement purification thereby offers significantly higher error thresholds for the required quality of local operations of the order of a few percent \cite{Du07}.

Here we show that a measurement-based implementation of entanglement purification allows for a remarkable increase of tolerability of errors by almost an order of magnitude. We derive a universal and optimal noise threshold for such measurement-based purification schemes, where we show that local white noise of up to 24\% per particle is tolerable for the resource states.

Entanglement purification allows one to obtain quantum states with a higher fidelity from several noisy copies in a probabilistic way by means of local operations. Recurrence protocols thereby operate on a small number of input pairs, leading to an increase of fidelity provided that the initial fidelity of the state is sufficiently high. A recursive application, i.e. taking the resulting pairs from a previous successful purification round as input for the next purification round, allows to eventually reach high-fidelity entangled states that can be used for the task in question. The achievable maximal fidelity as well as the required minimal fidelity of the pairs is thereby determined by the noise of local operations involved in the purification process, with a threshold value for noisy gates typical of the order of a few percent \cite{Du07}.

In \cite{Zw12} we have suggested a measurement-based implementation of entanglement purification protocols. Several copies of noisy entangled states are purified with help of locally prepared resource states. Rather than performing sequences of local two-qubit gates and single-particle measurements on the noisy pairs, one simply couples the particles via local Bell-measurements to the resource states. This accomplishes not only the implementation of the required gates, but at the same time allows one to determine from the outcome of the Bell measurements (with assistance of classical communication) whether the purification step was successful. The resource states are thereby of minimal size, i.e. for the purification of $m$ noisy pairs each local resource state has size $m+1$ - corresponding to the $m$ input particles plus one output particle. Notice that this approach may offer significant advantages as compared to gate-based approaches. In particular, no manipulation with coherent gates is required, but only a specific resource state needs to be generated and connected via Bell-measurements. The preparation of the resource state can be done beforehand and even in a probabilistic fashion. The performance and applicability of the measurement-based entanglement purification scheme is determined by the fidelity of the resource states. Depending on the physical set-up in question, e.g. in photonic systems, state preparation can be much easier or be done with higher accuracy than performing coherent two-qubit gates.

Here we report an additional advantage of the measurement-based approach, leading to a remarkable increase of the robustness of the protocols against local noise and imperfections. We show that subsequent purification rounds can be combined into a single step, where a $m+1$ particle state allows one to directly purify $m$ noisy Bell pairs \cite{Zw12}. Compared to a step-wise purification that involves several 3-particle entangled states, the total number of involved particles is reduced by a factor of almost one third. We show that this also leads to significantly relaxed requirement on the required fidelity of the resource states. While repetitive entanglement purification protocols operating on two copies of noisy entangled pairs at the same time (i.e. involving 3-particle resource states) have an error threshold of about $1-p=3.5\%$ of local white noise per particle, the $m+1$ particle resource state for the combined protocol accepts as much noise as $1-p=24\%$ per particle in the large $m$ limit, an additional robustness of almost one order of magnitude. While it is clear that multiparty states are more difficult to generate, leading to higher error rates and hence lower fidelities, we emphasize that here the acceptable noise {\em per particle} is enhanced. Even a constant noise $1-p$ per particle would lead to a required fidelity of the resource states that is exponentially small in system size, $F=\left(\frac{3p+1}{4}\right)^{m+1}$. Note that the threshold we find is optimal and universal. Universality follows from the independence of the results from a specific purification protocol, and optimality from the fact that only the (optimal) purification range of entanglement purification enters in the derivation.

\begin{figure}[htb]
\centering
\includegraphics[scale=0.3]{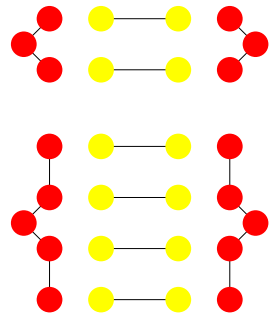}
%\put(-170,60){1}
\caption{(Color online) Resource states (red) up to local unitary operations for one (top) and two (bottom) purification steps using the protocol of Deutsch \emph{et al.}. We use standard graph state notation. The Bell pairs (yellow) are coupled to the resource states via Bell measurements. }
\label{FigResources}
\end{figure}

\begin{figure}[htb]
\centering
\includegraphics[scale=0.3]{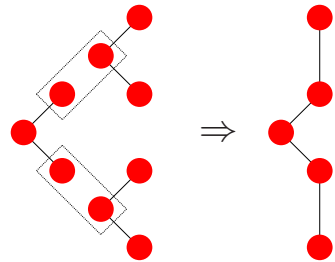}
\Large
\put(-50,45){$\Rightarrow$}
\caption{(Color online) Illustration of concatenation of resource states for two purification steps. The dashed boxes indicate Bell measurements.}
\label{FigConcatenation}
\end{figure}

\textit{Background and technical remarks.---}
We start by reviewing measurement-based entanglement purification. Measurement-based quantum computation \cite{Ra01,Br09} starts with a highly entangled resource state. Quantum circuits are translated to single qubit measurement patterns. There are several resource states which allow for universal quantum computation, e.g. the $2D$ cluster states \cite{Ra01b}. The read-in of a quantum state can be performed by joint Bell measurements on the input and the resource state. Clifford operations are implemented by Pauli measurements and can be done beforehand leading to smaller, special purpose resource states. Important examples for Clifford circuits are entanglement purification protocols and quantum error correcting codes. The idea of measurement-based entanglement purification is to implement the operations required in an entanglement purification protocol via measurement-based quantum computation. These protocols involve only Clifford gates and Pauli measurements and can be implemented with a resource state which requires only input and output qubits \cite{Ra03}. For example a measurement-based implementation of the protocol of Deutsch \emph{et al.} \cite{De96}, which maps two Bell pairs to one, requires two 3-qubit states. The resource states can be obtained in different ways. One possibility is to start with a sufficiently large $2D$ cluster state and the measurement pattern for the desired protocol and apply the transformation rules for graph states under Pauli measurements. Alternatively one can compute the associated Jamiolkowski state \cite{Ja72} of the desired map. The results of the Bell measurements at the read-in determine whether the purification step was successful and in addition a possible basis rotation of the resulting Bell pair. The resource states for one and two purification steps are shown in figure (\ref{FigResources}).
The concatenation of several purification steps can be done in two different ways. The output qubits of a resource state can be connected with the input qubits of the resource state of the subsequent purification step via a Bell measurement. This Bell measurement can be done beforehand since it can be decomposed into Clifford gates and Pauli measurements. One thus obtains a larger, single resource state, whereas the total number of involved qubits is reduced. This is illustrated in figure (\ref{FigConcatenation}). A second possibility is to compute the Jamiolkowski state of the whole map, consisting of several rounds of purification. It should be noted that the concatenation leads to a reduced overall success probability, since one has to require that all purification steps are successful at the same time, as discussed in \cite{Zw12}.

The main source of errors in measurement-based entanglement purification are imperfect resource states and imperfect Bell measurements. The latter can be incorporated in imperfect preparation of resource states and reduced fidelity of the Bell pairs. Hence the fidelity of the local resource states determines the purification interval, i.e. the minimal required and the maximal reachable fidelity. Here we model noise by local white noise (LWN). LWN can be seen as a worst case scenario, as any kind of local noise can be brought to this form []. Let $\rho=|\psi\rangle\langle \psi|$ denote the density matrix of a $n$ qubit resource state. We then model imperfect resource states by applying LWN on each of the particles,
\bea
\rho_{LWN}={\cal{D}}(p)\rho=\left(\prod_{j=1}^{n}{\cal{D}}^j(p)\right)\rho,
\eea
with
 \bea
 {\cal{D}}^j(p)\rho=p\rho + \frac{1-p}{2}\mathbb{I}^j\otimes \operatorname{tr}^j\rho,
 \eea
where $p\in[0,1]$ quantifies the level of noise. The Bell measurements at the read-in are assumed to be perfect, since imperfect measurements (e.g. modeled by white noise followed by perfect measurements) can be included in the LWN of resource states or Bell pairs, leading to a new, lower value of $p$.

It is important to note that one can exchange the location of LWN followed by a Bell measurement, i.e.
\bea
\label{eqLWN}
{\cal{P}}_{B}{\cal{D}}^1 \rho  = {\cal{P}}_{B}{\cal{D}}^2\rho,
\eea
where ${\cal{P}}_B\rho= P_B\rho P_B^{\dagger}$ and $P_B$ denotes a projector on a Bell state. We use the Jamiolkowski isomorphism which relates a completely positive map and a state, to proof this claim \cite{Su13a}. This result does not only hold for LWN but more generally for any noise of the form
\bea
\label{noise}
{\cal{M}}^j\rho=p\rho + (1-p)\sum_{i=0}^{3} \alpha_i \sigma^j_i \rho \sigma_i^{j\dagger}
\eea
with $\sum_{i=0}^{3}\alpha_i=1$ and Pauli operators $\sigma_0=\mathbb{I}, \sigma_1= X, \sigma_2 = Y, \sigma_3 = Z$. The superscript $j$ indicates the qubit on which ${\cal{M}}^j$ acts. The proof can be found in the supplemental material \cite{Su13a}.
The noise in (\ref{noise}) includes the examples of bit flip noise ${\cal{M}}_X(p)=\prod_{j=1}^{N}{\cal{M}}_X^j(p)$ with ${\cal{M}}_X^j\rho=p\rho + \frac{1-p}{2} \left(\rho+ X^j\rho X^j \right)$ as well as dephasing noise ${\cal{M}}_Z(p)=\prod_{j=1}^{N}{\cal{M}}_Z^j(p)$ with ${\cal{M}}_Z^j\rho=p\rho + \frac{1-p}{2} \left(\rho+ Z^j\rho Z^j \right)$.

%\section{Bell pairs}
\textit{Universal threshold for bipartite entanglement purification.---}
We investigate the thresholds for measurement-based, concatenated entanglement purification protocols for Bell pairs. In \cite{Zw12} it was observed that the error threshold for a measurement-based realization of the protocol by Deutsch \emph{et al.} \cite{De96} can be increased if the protocol is concatenated, leading to a new protocol which directly maps four pairs to one. Similarly one can concatenate several purification steps. Here we restrict to protocols which have a maximal purification interval: they start to purify when the fidelity of the noisy Bell pair exceeds one half and reach a fidelity of one in the limit of an infinite number of purification steps. This includes the protocols by Bennett \emph{et al.} \cite{Be96} and Deutsch \emph{et al.} \cite{De96}, as well as some entanglement purification protocols based on quantum error correction codes introduced by Aschauer in \cite{As05}. For the first two there exist analytical proofs for the purification interval \cite{Be96,Ma98} whereas for the latter ones the purification interval is determined numerically \cite{As05}.

The resource states which allow to implement these protocols will have only input and output qubits, since they only involve Clifford operations and Pauli measurements. The whole map is implemented by the Bell measurements at the read-in, there are no single qubit measurements in addition. Consequently we can move all noise from the resource state to the incoming and outgoing Bell pairs as described above and deal with a perfect purification. In the limit of infinitely many purification steps this means that the protocols output a perfect Bell pair, given that the initial fidelity, i.e. the overlap with the pure Bell state, of the noisy Bell pairs, after they are hit by the noise shifted from the resource state, is greater one half. The LWN of the resource state finally has to be applied to the resulting Bell pair.

This allows to determine the error thresholds for the protocols in the asymptotic limit in a quite simple way. Let us assume that there is LWN of $1-q$ on the Bell pairs and LWN of $1-p$ on the resource state. The fidelity of the Bell pairs to be purified is then $F_1=\frac{1+3q^2}{4}$, whereas the fidelity of the output pairs is $F_2=\frac{1+3p^2}{4}$, since one has to apply the LWN of the resource state. The purification protocols require that the fidelity of the incoming Bell pairs, after they are hit by the additional noise of the resource state, is larger than one half. Thus one has
\bea
\label{ineq1}
\bra{\phi^{+}}{\cal{D}}(q){\cal{D}}(p)\rho_{\phi^{+}} \ket{\phi^{+}}= \frac{1+3p^2q^2}{4} > \frac{1}{2}.
\eea
In addition the fidelity of the output pairs $F_2$ has to be larger than the fidelity of the input pairs $F_1$,
\bea
\label{ineq2}
 \frac{1+3p^2}{4} > \frac{1+3q^2}{4}.
 \eea
 From (\ref{ineq2}) it follows that $p > q$, whereas from (\ref{ineq1}) we get $pq > \frac{1}{\sqrt{3}}$. The threshold value of $p$, i.e. the value $p_{min}$ such that purification is possible for $p > p_{min}$ is then given by $p_{min}=\frac{1}{\sqrt[4]{3}}$. This means that in the asymptotic limit these protocols can tolerate a noise level of $1-p_{min}=1-\frac{1}{\sqrt[4]{3}} \approx 24.0\%$, before they break down.
 At no point in the derivation of this result we make use of the specifics of the protocols such as the gates and measurements, the only thing that matters is the maximal purification interval. This implies that --within the used error model-- the threshold we find is \emph{optimal} and \emph{universal}.

In addition we analyze numerically the error thresholds for a finite number of concatenations of the protocol of Deutsch \emph{et al.} and a protocol based on the $[[ 5,1,3 ]]$ quantum error correcting code \cite{As05}. The results are shown in tables \ref{taboxford} and \ref{tabaschauer}. Already a small number of concatenations suffices to push the error thresholds for both protocols beyond 20\%. This can be of practical advantage, e.g. for measurement-based quantum repeaters \cite{Zw12}.

\begin{table}[htb]
\caption{Error thresholds $1-p$ for various numbers of concatenations of the protocol of Deutsch \emph{et al.}. The second column shows the number of input pairs that are mapped to the number of output pairs in one step.}
\vspace{0.5cm}
\centering
\begin{tabular}{c | c | c}
\hline\hline
\# concatenations & mapping & threshold in \% \\
\hline
$0$ & $2\rightarrow1$ & $3.5$\\
\hline
$1$ & $4\rightarrow1$ & $7.1$\\
\hline
$2$ & $8\rightarrow1$ & $10.4$\\
\hline
$4$ & $32\rightarrow1$ & $15.4$\\
\hline
$7$ & $256\rightarrow1$ & $20.1$\\
\hline
\end{tabular}
\label{taboxford}
\end{table}

\begin{table}[htb]
\caption{Error thresholds $1-p$ for various numbers of concatenations of the $[[5,1,3]]$ protocol. The second column shows the number of input pairs that are mapped to the number of output pairs in one step.}
\vspace{0.5cm}
\centering
\begin{tabular}{c | c | c}
\hline\hline
\# concatenations & mapping & threshold in \% \\
\hline
$0$ & $5\rightarrow1$ & $6.7$\\
\hline
$1$ & $25\rightarrow1$ & $13.3$\\
\hline
$2$ & $125\rightarrow1$ & $17.3$\\
\hline
$3$ & $625\rightarrow1$ & $20.2$\\
\hline
\end{tabular}
\label{tabaschauer}
\end{table}

The maximally reachable fidelities for some protocols and different values of $1-p$ are shown in tables \ref{taboxfordfid} and \ref{tabaschauerfid}. 

\begin{table}[htb]
\caption{Reachable fidelity for selected protocols based on the protocol of Deutsch \emph{et al.} for different noise levels $1-p$.}
\vspace{0.5cm}
\centering
\begin{tabular}{c || c | c | c | c}
\hline\hline
 protocol & $1-p=1\%$ & $1-p=3\%$ & $1-p=5\%$ & $1-p=10\%$ \\
\hline\hline
$2\rightarrow1$ & $96.2\%$ & $84.7\%$ & n/a & n/a\\
\hline
$4\rightarrow1$ & $98.4\%$ & $94.7\%$ & $89.5\%$ & n/a\\
\hline
$8\rightarrow1$ & $98.5\%$ & $95.5\%$ & $92.4\%$ & $80.4\%$\\
\hline
$32\rightarrow1$ & $98.5\%$ & $95.6\%$ & $92.7\%$ & $85.7\%$\\
\hline
$256\rightarrow1$ & $98.5\%$ & $95.6\%$ & $92.7\%$ & $85.8\%$\\
\hline
\end{tabular}
\label{taboxfordfid}
\end{table}

\begin{table}[htb]
\caption{Reachable fidelity for selected protocols based on the $[[5,1,3]]$ protocol for different noise levels $1-p$.}
\vspace{0.5cm}
\centering
\begin{tabular}{c || c | c | c | c}
\hline\hline
 protocol & $1-p=1\%$ & $1-p=3\%$ & $1-p=5\%$ & $1-p=10\%$ \\
\hline\hline
$5\rightarrow1$ & $97.5\%$ & $92.1\%$ & $85.6\%$ & n/a\\
\hline
$25\rightarrow1$ & $98.5\%$ & $95.6\%$ & $92.6\%$ & $84.7\%$\\
\hline
$125\rightarrow1$ & $98.5\%$ & $95.6\%$ & $92.7\%$ & $85.7\%$\\
\hline
$625\rightarrow1$ & $98.5\%$ & $95.6\%$ & $92.7\%$ & $85.8\%$\\
\hline
\end{tabular}
\label{tabaschauerfid}
\end{table}

\textit{Entanglement purification of multipartite graph states.---}
In this chapter we investigate error thresholds for purification protocols for  two-colorable graph states, which includes cluster states and GHZ states. Two-colorable graph states are equivalent to codewords of CSS quantum error correcting codes \cite{Lo04}. The protocols were introduced in \cite{Du03} and only involve Clifford gates and Pauli measurements. Consequently they also allow for a very compact, measurement-based implementation using resource states with qubits for input and output only. One can then use the same techniques as above to determine the error thresholds in the limit of an infinite number of purification steps.

Let $\rho_G$ denote the density matrix of a graph state $\ket{G}$ and $q_{min}$ the minimal value of $q$ such that a noisy graph state ${\cal{D}}(q)\rho_G$ can be purified. The maximal fidelity that can be reached by using a noisy resource state implementing infinitely many purification steps with LWN parameter $p$ is $\bra{G} {\cal{D}}(p)\rho_G \ket{G}$. This simply follows from the fact that one can move the noise on the resource state (characterized by $p$) which is used to perform the purification, to the finally resulting graph state. In order to have purification, the reachable fidelity has to be larger than the fidelity of the state ${\cal{D}}(q)\rho_G$ that shall be purified. Furthermore, the fidelity of the input state, after moving the additional noise of the resource state to it, has to be larger or equal than the minimal fidelity required for purification:
\bea
\bra{G} {\cal{D}}(p){\cal{D}}(q)\rho_G \ket{G} \geq \bra{G} {\cal{D}}(q_{min})\rho_G \ket{G}.
\eea
This immediately gives $qp \geq q_{min}$ together with $p > q$. The threshold value $p_{min}$ such that purification is possible for $p>p_{min}$ is then given by $p_{min}=\sqrt{q_{min}}$.

In \cite{As05b} the values for $q_{min}$ have been determined numerically for linear cluster states and GHZ states up to size $N=10$. For the $10$ qubit linear cluster state $q_{min} \approx 0.6$, leading to $p_{min} \approx 0.77$. This means that the measurement-based concatenated protocol can tolerate $1-p\approx 23\%$ noise before it breaks down. The purification protocol for a $N=10$ qubit GHZ state can tolerate a value of $q_{min}\approx0.8$. The reason that it is much higher than for the linear cluster state is the larger vertex degree. The measurement-based protocol can then tolerate up to $1-p=1-\sqrt{q_{min}}\approx 11\%$ noise in the limit of infinitely many purification steps.

In addition, in \cite{As05b} the error thresholds for bit flip noise ${\cal{M}}_X(q)$ have been investigated. Closed linear cluster states can be purified as long as $q \geq q_{min}\approx0.4938$, which in the measurement-based approach translates to $1-p\approx29.7\%$ bit flip noise on the resource state in the asymptotic limit.

\textit{Summary and outlook.---}
We have investigated measurement-based entanglement purification protocols in the presence of noise and imperfections. In addition to possible principal advantages of such a measurement-based approach, we encountered a significant increase of tolerable errors for local resource states. We derived a universal, protocol independent error threshold, where we find that the acceptable noise per particle can be as high as 24\%. The reason for the increased stability is unique to the measurement-based approach, and is based on the reduction of the size of the resource states when considering several rounds of entanglement purification. A similar result holds for the measurement-based purification of graph states.

Given the importance of entanglement purification as a tool for the generation of distributed high fidelity entanglement, the encountered stability against local imperfections may open the way towards a practical implementation of the proposed schemes. Similar ideas can be applied e.g. in the context of measurement-based quantum error correction (see \cite{Zw13}).

\textit{Acknowledgements.---}
This work was supported by the Austrian Science Fund (FWF): P24273-N16, SFB F40-FoQus F4012-N16.

\section{Supplemental Material}

\textit{Proof of equation (\ref{eqLWN}). ---} Let $M^i$ denote the map $\rho \rightarrow {\cal{P}}_{B}^{12}{\cal{D}}^i \rho $ for $i \in \{1,2\}$. Then
\bea
M^1\left( \rho_{\phi^{+}}^{13}\otimes \rho_{\phi^{+}}^{24} \right) & = & {\cal{P}}_{B}^{12}{\cal{D}}^1\rho_{\phi^{+}}^{13}\otimes \rho_{\phi^{+}}^{24}  \nonumber \\
& = & {\cal{P}}_{B}^{12}{\cal{D}}^3\rho_{\phi^{+}}^{13}\otimes \rho_{\phi^{+}}^{24}  \nonumber \\
& =& \rho_{B}^{12} \otimes {\cal{D}}^3 \rho_{B}^{34} \nonumber \\
& = & \rho_{B}^{12} \otimes {\cal{D}}^4 \rho_{B}^{34} \nonumber \\
& = & {\cal{P}}_{B}^{12} \rho_{\phi^{+}}^{13}\otimes {\cal{D}}^4\rho_{\phi^{+}}^{24} \nonumber \\
& = & {\cal{P}}_{B}^{12} \rho_{\phi^{+}}^{13}\otimes {\cal{D}}^2\rho_{\phi^{+}}^{24} \nonumber \\
& = & M^2\left( \rho_{\phi^{+}}^{13}\otimes \rho_{\phi^{+}}^{24} \right), %\nonumber
\eea
where we use that ${\cal{D}}^1 \rho_{B}={\cal{D}}^2 \rho_{B}$. Since the two Jamiolkowski states $M^1\left( \rho_{\phi^{+}}^{13}\otimes \rho_{\phi^{+}}^{24} \right)$ and $M^2\left( \rho_{\phi^{+}}^{13}\otimes \rho_{\phi^{+}}^{24} \right)$ are identical the two maps $M^1$ and $M^2$ are the same.

\textit{Proof of equation (\ref{noise}). ---} The proof is similar to the one above, it suffices to show that ${\cal{M}}^1 \rho_{B}={\cal{M}}^2 \rho_{B}$. Let $\rho_{B,k}=\sigma_k^1 \rho_{\phi^+} \sigma_k^1$ for $k \in \{0,1,2,3\}$. Then
\bea
{\cal{M}}^1\rho_{B,k} & = & p\rho_{B,k}+(1-p)\sum_{i=0}^{3} \alpha_i \sigma^1_i\rho_{B}\sigma_i^{1\dagger} \nonumber \\
& = & p\rho_{B,k}+(1-p)\sum_{i=0}^{3} \alpha_i \sigma^1_i\sigma_k^1 \rho_{\phi^+} \sigma_k^1 \sigma_i^{1\dagger} \nonumber \\
& = & p\rho_{B,k}+(1-p)\sum_{i=0}^{3} \alpha_i \sigma^1_k\sigma_i^1 \rho_{\phi^+} \sigma_i^1 \sigma_k^{1\dagger} \nonumber \\
& = & p\rho_{B,k}+(1-p)\sum_{i=0}^{3} \alpha_i \sigma^1_k\sigma_i^2 \rho_{\phi^+} \sigma_i^2 \sigma_k^{1\dagger} \nonumber \\
& = & p\rho_{B,k}+(1-p)\sum_{i=0}^{3} \alpha_i \sigma^2_i\sigma_k^1 \rho_{\phi^+} \sigma_k^1 \sigma_i^{2\dagger} \nonumber \\
& = & {\cal{M}}^2\rho_{B,k}
\eea
where we use $\{\sigma_i,\sigma_j\}=2\delta_{ij}\mathbb{I}$.

\end{document}